# Ferromagnetic Semiconductors and Spintronic Devices

Nazmul Hasan, *Graduate Student Member, IEEE*

*Abstract*— Ferromagnetic semiconductors play a crucial role in spintronic devices, enabling effective control of electron spin over charge. This study explores their unique properties, ongoing advancements in spin control, and potential integration into next-generation semiconductor technologies.

## I. INTRODUCTION

DISCOVERY of giant magnetoresistance (GMR) and tunneling magnetoresistance (TMR) during 1980s-2000s opened a new avenue in solid state physics to understand and control both charge transport and spin behavior of carriers in semiconductors [1]. However, semiconductors' physical transport behavior is extraordinarily sensitive to defects, impurities, gate biases, and so on whereas magnetism comes from an ordered state of collective electronic formations which have a significant impact on material's optoelectronic properties. Introducing local moments into typical semiconductors in 1970s led to the realization that it is possible to tune materials' physics by combining semiconducting and magnetic behavior through diluted magnetic semiconductors [2].

However, it has established that temperature dependent ferromagnetic transition is possible into semiconductor materials when III-V materials are heavily doped with Mn utilizing valence band (VB) carriers. Strong interaction of VB charge carriers of s shell with localized magnetic moments of partially filled d/f shell significantly affect magnetic order and thus carrier motion [3]. Besides, applying external magnetic field also can alter semiconductors' optoelectronic behavior through creating or manipulating the flow of spin-polarized carriers [4]. Hence, manipulating materials' properties utilizing their both semiconducting and magnetic behavior through controlling carrier concentrations and magnetic order by applying external magnetic field is become intriguing in the solid-state physics for next generation electronics devices. Spintronic devices utilize this spin-current transport behavior following various device engineering techniques for high-speed information processing.

To understand the functional ferromagnetic semiconductor spintronics, initiation physics of ferromagnetism in semiconductors described with spintronics device principles in the following section 2 and 3. Moreover, recent device engineering techniques from materials to scaling are also shortly drew up in section 4, and at last section practical applications perspectives are discussed.

## II. ORIGIN OF FERROMAGNETISM IN SEMICONDUCTORS

In semiconductor, universally, interplay between electronic spin degree of freedom, repulsive coulomb interactions between electrons, and the fermionic quantum statistics of electrons arises ferromagnetism in materials where Pauli exclusion principle play a vital role to correlate the spin and orbital behavior to determine particle exchange symmetry in electronic wave functions [1], [5]. The repulsive coulomb interactions between electrons strongly influence magnetic order of the material and aligning with Pauli's principle electrons' groups sharing same spin state results to having quantum ground state of the system. In ferromagnetic case, a nonzero local spin density of states is associated with the many-electron wave function of the system in the same direction of space at ground state [3], [6]. This dependence of the system energy on the orientation order of magnetic ion's local moments is called exchange interaction which can be controlled by doping in the host semiconductor material. Hence, fermi statistics explain the origin of ferromagnetic behavior in semiconductors as illustrated in Fig. 1 below. Magnetic semiconductors usually have narrower carrier energy bands where s orbital electrons strongly interact with partially filled d/f shell's magnetic moments.

Considering Mn doped GaAs semiconductor, where the system energy is dependent on the exchange interaction of Mn based on its local moments' orientation. Intrinsically, itinerant exchange of electron spins can result in spontaneous spin polarization of the entire system as result of having same spin state. Which will lead to double occupancy in each eigen states with a large density of states at fermi level that in turn makes possible to move electrons from one spin band to another having system's kinetic energy lower. Hence, Stoner instability happens in the ferromagnetic system, and thus Stoner mechanism of localized key spin failed in ferromagnetism [4]. However, Heisenberg's direct exchange in the local nature of two spin moments in effect of strong local coulomb interactions difference of singlet spin wave and triplet spin wave suppress valence charges fluctuations. That leads to transferring nonmagnetic atom's electron to the favorable empty shell of magnetic atoms resulting in polarized nonmagnetic atom by electrons local moment that is coupled with neighboring magnetic atoms as shown in Fig. 1(b). This exchange coupled magnetic moments form parallel spin alignment ferromagnetically increasing hopping probability by decreasing kinetic energy. Hence, conduction electrons tend to maintain ferromagnetically favored system as the electron energy is minimal though the ordering is dependent on the curie temperature Tc of the system [7].

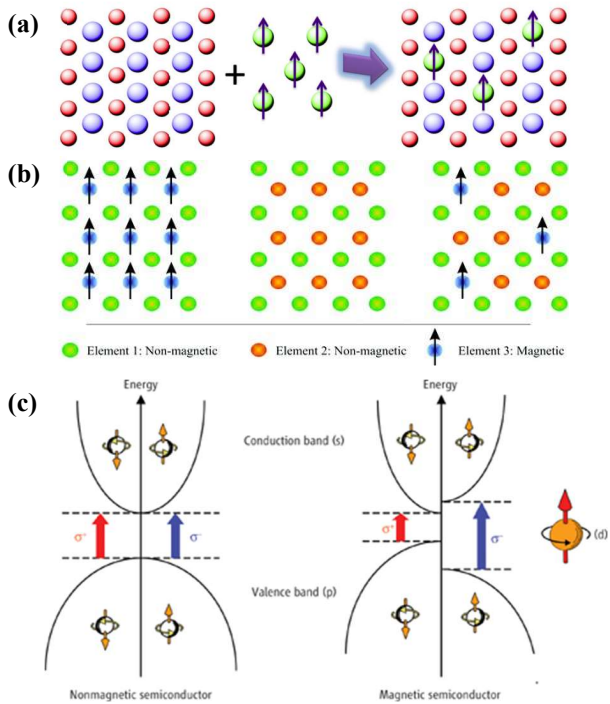

**Fig. 1.** (a) & (b) Magnetic behavior mechanism for Ferromagnetic materials, (c) Changes in band structure based on spin orientation [8].

III. SEMICONDUCTOR SPINTRONIC DEVICE PRINCIPLES

Adding spin degree of freedom in electrons has ushered in a new era of versatility, allowing for precise control over the magnetic spin behavior of electrons, particularly in the realm of novel nanoelectronic devices. Semiconductor spintronics has introduced crucial concepts in spin transport, spin injection, Silsbee-Johnson spin-charge coupling, spin-dependent tunneling, spin relaxation, and spin dynamics. This intricate interplay between varying symmetries and structures results in diverse forms of spin-orbit coupling within the semiconductor-based heterostructures system, yielding a spectrum of effective spin-orbit Hamiltonians that provides a versatile and adaptable framework for manipulating and controlling spin-related phenomena in semiconductor materials [6].

Alike semiconductor doping concentration led quantum tunneling mechanism in ferromagnetic semiconductors also holds pivotal significance for the field of spintronics particularly in comprehending the transport of spin-polarized electrons through materials. Quantum tunneling is observed in ferromagnetic semiconductors devices, particularly in magnetic tunnel junctions (MTJs) where two ferromagnetic layers are separated by a thin insulating barrier that allows electrons to traverse potential energy barriers possessing their specific spin orientations resulting in a controlled flow of spin-polarized electrons [9]. According to quantum mechanics, there exists a finite probability that these electrons can tunnel through the barrier despite its energy being higher than the electrons' energy. The probability of tunneling is influenced by the spin alignment of the electrons and the magnetic configuration of the ferromagnetic layers. When the magnetizations of the layers are parallel, electrons with aligned spins face a lower tunneling barrier, leading to a higher probability of tunneling which results in a spin-polarized current. Utilizing tunneling magnetoresistance (TMR) in devices leveraging this phenomenon for efficient manipulation and detection of spin-polarized electrons through controlled resistance based on the relative alignment of magnetization in ferromagnetic layers [6].

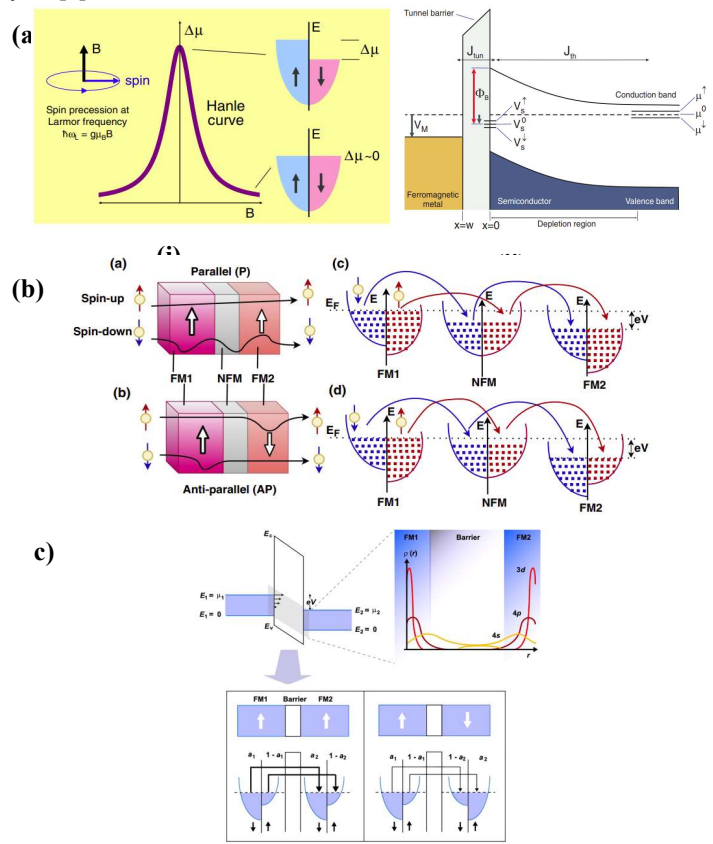

**Fig. 2.** (a) (i) Hanle effect results to manipulation of the spin-polarization in semiconductor, (ii) Energy band profile of FM/I/SC junction [10]; (b) Spin valve based GMR effect: parallel orientation facilitating easy flow for majority electrons (spin-up), antiparallel direction leads in a scattered flow of electrons projected with DOS [9]; (c) Spin-orientation based electron tunneling in FM heterostructure [11].

The quantum tunneling mechanism is involved in spin-transfer torque, a phenomenon where a spin-polarized current exerts torque on a magnetic layer enables the precise control of magnetization direction in spintronics devices like magnetic random-access memory (MRAM) [12]. Maintaining quantum coherence during tunneling spin polarized electron transport ensures the integrity of spin information. The ability to selectively transmit spin-polarized electrons through barriers allows for the efficient operation of spintronic components [13]. Quantum tunneling enables the creation of devices with reduced power consumption and enhanced performance. As spintronics explores avenues for quantum computing, the quantum tunneling mechanism in ferromagnetic semiconductors becomes a critical element for the creation of qubits and the manipulation of quantum states, offering potential advantages in quantum information processing [13], [14].

In ferromagnetic semiconductor-based spintronics devices, the impact of spin-orbit coupling (SOC) is deeply rooted in the fundamental interplay between the intrinsic spin of electrons and their orbital motion. This interaction can be described by the Rashba and Dresselhaus effects in which structural inversion asymmetry, such as in asymmetric semiconductor heterostructures results to the Rashba effect, while the Dresselhaus effect arises from bulk inversion asymmetry, common in materials lacking structural inversion symmetry [15]. Spin-orbit coupling modifies the energy levels of electrons depending on their spin orientation, effectively mixing different spin states. This coupling is a relativistic effect that becomes particularly significant in ferromagnetic semiconductor devices, instrumental in spin manipulation, notably through the spin-orbit torque phenomenon.

## IV. FUNCTIONAL MATERIALS AND DEVICES FABRICATION

Semiconductor properties can be manipulated through processes such as doping (by density and type), exposure to light, and modulation by electrostatic gates which enable amplification and transistor-like functionalities. Advancement in materials and device engineering, allowing for tailored functionalities and improved performance in ferromagnetic spintronic devices. The design and fabrication of such devices involve integrating different materials to create layered configurations that exploit the unique properties of each component. For instance, combining ferromagnetic metals with semiconductors or insulators enables the creation of spintronic devices with enhanced spin transport and manipulation capabilities. Leveraging bandgap engineering, diverse heterostructures can be constructed to tune ferromagnetic semiconductors utilizing the density of states in spin-polarized energy bands. Recent studies also attempted to utilize the van der walls materials to make heterostructures in view of implementing next generation computing electronics enhancing spin electron transport, resistance, switching behavior, spin injection modulation and so on.

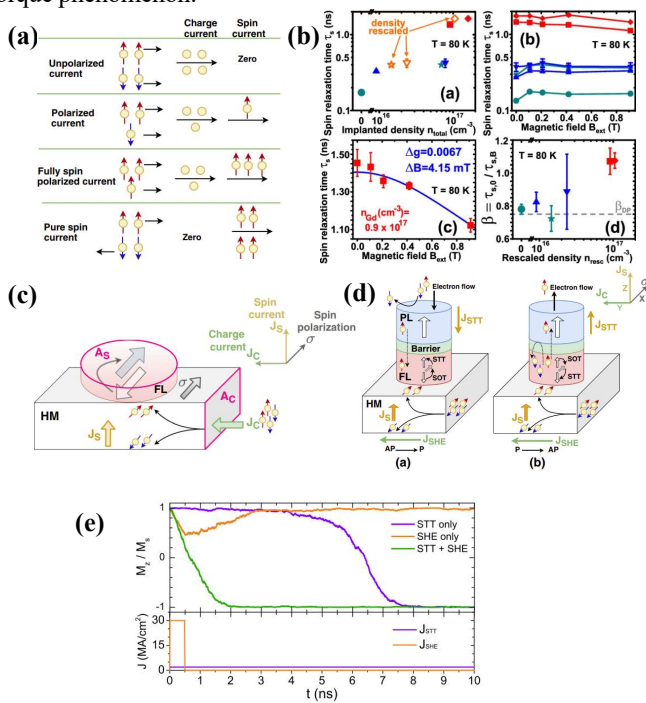

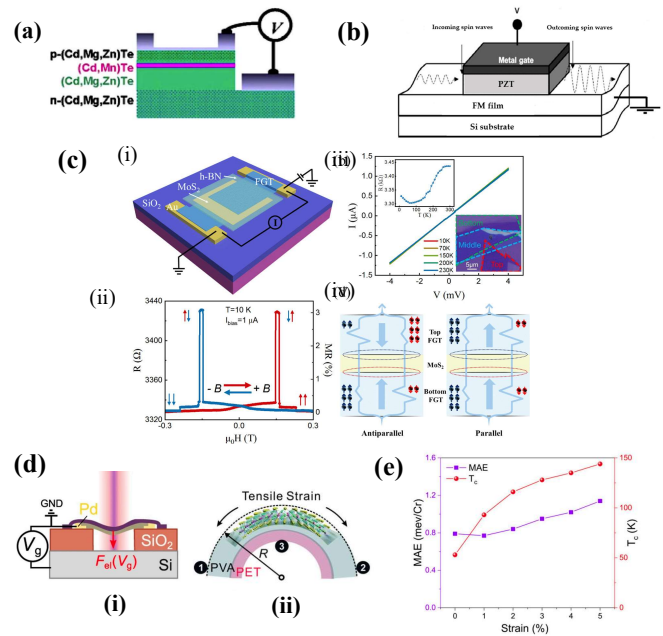

**Fig. 3.** (a) Electron and Spin current behavior with polarizations [16] (b) Spin relaxation time ss vs. implanted density, spin relaxation time behavior on magnetic field [17], (c) charge-spin current conversion and (d) switching mechanism in FM devices, (e) Magnetization behavior with applied current pulse for different switching [9].

**Fig. 4.** Device configurations for (a) Heterostructure (b) layered structure combined with ME/PE; (c)(i)Device Schematics of the FGT/MoS$_2$/FGT spin valve with h-BN capping layer, (ii) Junction resistance and MR vs magnetic field, (iii) I-V curve for the device, (iv) spin transport w.r.t. spin valve is in antiparallel (left) and parallel (right) configurations [18]; (d) electrostatic strain methods; (e) MAE per atom and T$_C$ for strain for monolayer FM material [19].

Employing an interacting device structure utilizing interaction of charge carriers and spin waves can result in spin wave amplification by charge carriers drifting mobility [7]. In case of spin wave modulation, mutual orientation between magnetic and electric field is important to maintain in the device, where device and materials engineering like use of ferromagnetic thin film with piezoelectric layer can modify the device performance. Another most popular technique is followed by device researchers is spin valve structure, that a device based on ferromagnetic semiconductor thin film heterostructures that utilizes the Giant Magnetoresistive effect by tunneling spin-polarized charge carriers [11]. Moreover, strain engineering is becoming an intriguing, applied approach to tune the ferromagnetic behavior in semiconductor heterostructures in a more controllable and precise way to advance spintronics applications particularly using 2d van der walls materials [14], [20].

## IV. Applications

Ferromagnetic semiconductors, with their unique electronic and magnetic properties, find compelling applications in the realm of spintronic device physics. In memory and logic applications, the integration of ferromagnetic semiconductors enables the creation of spintronic devices that capitalize on the spin degree of freedom of electrons. Non-volatile magnetic memory, exemplified by Magnetoresistive Random Access Memory (MRAM), leverages the ferromagnetic properties for stable data storage with advantages such as low power consumption and high-speed operation [8]. The utilization of ferromagnetic semiconductors in Quantum dots enhances spintronic functionalities, contributing to the development of innovative spin-based computing and data storage technologies [3]. The quantum nature of these materials allows for the manipulation and storage of quantum information. In sensing applications, ferromagnetic semiconductors play a crucial role, particularly in magnetoresistive sensors, where changes in resistance due to magnetic fields enable sensitive and precise detection. Spintronic detectors, utilizing the properties of ferromagnetic semiconductors, demonstrate efficacy in sensing magnetic fields and currents, presenting opportunities for applications across various industries, including automotive, healthcare, and electronics, where reliable and high-performance sensing and detection are imperative for technological advancement [6]. The versatile physics of ferromagnetic semiconductors thus underpin their indispensable role in the ever-evolving landscape of spintronics.

## V. Conclusion

Taking advantage of having the ability to tune both electron and spin together in the ferromagnetic semiconductor, researchers are emphasizing their efforts to come up with the most efficient computing devices. Tunability by core semiconductor variable to spin transport paved the way to scaling the devices for next generation spintronics utilizing the spin degree of freedom effect in device performances. Thus, understanding the device mechanism with controlled spin polarized electrons' flow requires more efforts through materials and device engineering. Adhering to Moore's Law, achieving efficiency in nano-scaled devices requires mastery over both electron and spin behavior in ferromagnetic semiconductor spintronic devices.

## Acknowledgment

The author appreciates Professor Roman Sobolewski for guiding and reviewing the term paper summary, enhancing the manuscript.